\documentclass[aps,twocolumn, secnumarabic,nobalancelastpage,amsmath,amssymb,nofootinbib]{revtex4}
\usepackage{array}
\usepackage{graphicx}
\usepackage{bm}% bold math
\usepackage{bbm}
\usepackage{bbold}
\usepackage{amsmath}
\usepackage{braket}
\usepackage{color}
\usepackage{ulem}
\usepackage{tabularx}

\usepackage{todonotes}
\usepackage{xcolor}

\usepackage[colorlinks=true, linkcolor=blue, citecolor=blue, urlcolor=blue]{hyperref}

\usepackage{pdfrender}  % more text control
\usepackage[percent]{overpic}  % text over figures
\usepackage{ifthen}
\newcommand{\gfxext}{pdf}  % pdf or png
\newcommand{\extrabold}[1]{%
  {\pdfrender{TextRenderingMode=FillStroke,LineWidth=0.10pt}\textbf{#1}}%
}
\newcommand{\captionFont}[1]{%
    \extrabold{#1}%
}

\newcommand{\figCap}[2][true]{%
    \ifthenelse{%
        \equal{#1}{true}}
        {(}%
        {}%
    \captionFont{#2}%
    \ifthenelse{%
        \equal{#1}{true}}
        {)}%
        {}%
}

\begin{document}

\title{Single- and Two-Mode Squeezing by Modulated Coupling to a Rabi Driven Qubit}

\author{E. Blumenthal$^{_1}$}
\author{N. Gutman$^{_2}$}
\author{I. Kaminer$^{_2}$}
\author{S. Hacohen-Gourgy$^{_1}$} 

\affiliation{$^{_1}$Department of Physics, Technion - Israel Institute of Technology, Haifa 32000, Israel
\\
$^{_2}$Department of Electrical and Computer Engineering, Technion - Israel Institute of Technology, Haifa 32000, Israel
}

\begin{abstract}
    Advanced bosonic quantum computing architectures demand nonlocal Gaussian operations such as two-mode squeezing to unlock universal control, enable entanglement generation, and implement logical operations across distributed modes.
This work presents a novel method for generating conditional squeezing using a Rabi-driven qubit dispersively coupled to one or two harmonic oscillators. A proof that this enables universal control over bosonic modes is provided, expanding the toolkit for continuous-variable quantum information processing. Using modulated Jaynes-Cummings interactions in circuit QED, the simulation predicts intra-cavity squeezing of 13dB (single-mode), 4dB (superimposed single-mode), and 12dB (two-mode), with the latter two yet to be demonstrated experimentally. These results establish a new paradigm for qubit-conditioned control of photonic states, with applications to quantum sensing and continuous-variable computation on readily available systems.
\end{abstract}

\maketitle

\section{Introduction}
Quantum squeezed states and quantum squeezing operations are widely used for quantum sensing~\cite{pan2024realisation, zheng2016accelerating}. They are useful for improving the signal-to-noise ratio by amplifying one quadrature and squeezing the other. Amplification may overwhelm noise in an amplification chain~\cite{Krantz2019Quantum}, and squeezing can reduce the quantum uncertainty of an observable~\cite{Aasi2013}. Quantum squeezed states are usually generated by a designated device, such as a Josephson parametric amplifier at microwave frequencies~\cite{Zhong2013Squeezing} or spontaneous parametric down-conversion crystals at optical wavelengths~\cite{Slusher1985bservation}. As squeezed states are sensitive to photon loss, it is advantageous to minimize their transportation distance by generating them inside the observed system. Such intra-cavity single mode squeezing was recently achieved in a circuit QED setup by using gated conditional displacements ~\cite{Eickbusch2022, Hastrup2021Unconditional} or by utilizing Kerr non-linearity~\cite{Cai2025Quantum}. It can be used for sensitive detection of displacements of the cavity electromagnetic field or for quantum computation using continuous variable encoding~\cite{Copetudo2024Shaping}. For example, squeezing was demonstrated to extend the lifetime of a cat-state inside a superconducting cavity~\cite{pan2022protecting} and was proposed to enable generation of Gottesman-Kitaev-Preskill states~\cite{Hastrup2022Protocol}.

Recent advances in quantum science and technology increasingly demand operations that go beyond single-mode squeezing. Applications ranging from quantum error correction to quantum simulation and sensing now call for more complex forms of control—such as two-mode squeezing, which entangles bosonic modes, and conditional operations that couple discrete and continuous-variable degrees of freedom. These higher-dimensional squeezing operations promise to unlock sophisticated protocols that enhance the robustness and scalability of continuous-variable quantum architectures.

In this letter, we propose a scheme for generation of continuous single- and two-mode squeezing (TMS) using a Rabi-driven qubit that is dispersively coupled to two quantum harmonic oscillators. Our operation is conditional, since the squeezed axis is conditioned on the state of the qubit. We will show how our scheme can be realized in a circuit QED setup and present results of numerical simulations. Our results predict generation of intra-cavity squeezed state with $13~\mathrm{dB}$ of squeezing for single-mode squeezing and $12~\mathrm{dB}$ of squeezing for two-mode squeezing, which is in line with the largest amplitude reported in experiment to date for single-mode squeezing~\cite{Eickbusch2022, Cai2025Quantum}. For a superposition of squeezed states, we achieve $~4~\mathrm{dB}$ of conditional squeezing in the simulation. The squeezed amplitude of a superposition of squeezed states is currently limited due to high-order effects of our scheme.

Conditional-squeezing could be used for control of a rotation-symmetric Bosonic code~\cite{Grimsmo2020Quantum} with even and odd superpositions of squeezed states as the computational basis~\cite{hope2025preparation}. We provide a proof of universal control using this operation together with single qubit rotations and displacements. 

During the preparation of this manuscript, we learned of a recent independent proposal suggesting conditional- (or controlled-) squeezing on a weekly anharmonic oscillator that is dispersively coupled to a transmon qubit~\cite{delgrosso2025controlled}. In their proposal, a SQUID is required for a two-photon drive. Our study extends the notion of conditional-squeezing to two-modes, without the need for extra hardware.
\begin{figure}[ht]
    \centering
    \vspace{0.5em}
    \begin{overpic}[width=0.95\linewidth]{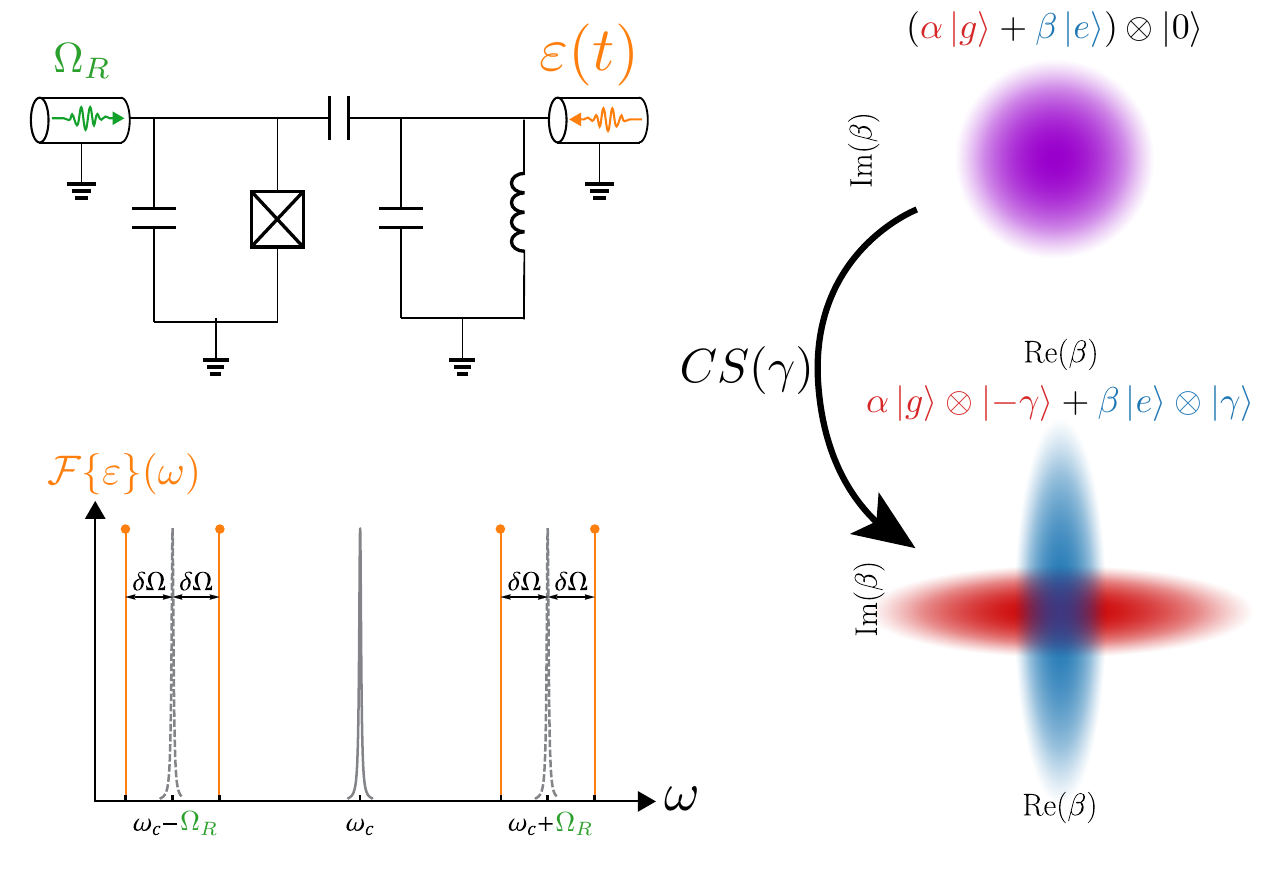}
        \put(-1,70){\figCap[0]{a}}
        \put(-1,35){\figCap[0]{b}}
        \put(60,70){\figCap[0]{c}}
    \end{overpic}
    \caption{%
        \textbf{Scheme Overview.} %
        \figCap{a} A circuit QED system, composed of a Rabi-driven qubit, and a sideband-driven cavity mode (represented by an LC circuit).  
        \figCap{b} Frequency representation of the 4 sidebands drive around the resonance frequency of the cavity mode.  
        \figCap{c} A qualitative representation of a Wigner function before and after application of the conditional-squeezing operation.
    }
    \label{fig:schemeOverview}
\end{figure}

\section{Theory}
\subsection{Single-mode squeezing Hamiltonian}
Let us consider a system composed of a qubit and a quantum harmonic oscillator coupled by a time-modulated interaction described by
\begin{equation}\label{eq:Hcoupling}
\begin{split}
        H_\mathrm{coup} &=\mathrm{i}\cos(\delta\Omega t)\left(g\sigma_+a - g^*\sigma_-a^\dagger\right) \\
        &-\sin(\delta\Omega t) \left(g\sigma_+a^\dagger+g^*\sigma_-a\right),
\end{split}
\end{equation}
in the interaction picture. Here $\sigma_+$ and $\sigma_-$ are the raising and lowering operators of the qubit, $a^\dagger$ and $a$ are the creation and annihilation operators of excitations in the harmonic oscillator, $g$ is the coupling strength and $\delta\Omega$ is the modulation frequency. This coupling contains a Jeynes-Cummings (JC)~\cite{Jaynes1963Comparison} term
modulated by $\cos(\delta\Omega t)$ and an anti-JC term that is modulated by $\sin(\delta\Omega t)$. We assume that $\delta\Omega\gg g$, so that we can use the Magnus expansion~\cite{Blanes2009} to expand the Hamiltonian in orders of $g/\delta\Omega$. The approximated mean Hamiltonian integrated over one period of $2\pi/\delta\Omega$ is (For all calculations, see~\cite{git})
\begin{equation}\label{eq:Hsqueezing}
    H_\mathrm{squeezing} = \frac{1}{2\delta\Omega}\sigma_z\left(\left(g^*\right)^2 a a + g^2 a^\dagger a^\dagger \right) + \mathcal{O}\left(\left(\frac{g}{\delta\Omega}\right)^2\right).
\end{equation}
This effective Hamiltonian describes a conditional squeezing interaction at a rate of $g^2/2\delta\Omega$. We will detail how the modulated coupling of Equation~\ref{eq:Hcoupling} can be generated on a circuit QED system to achieve the conditional squeezing interaction of Equation~\ref{eq:Hsqueezing}. This is accomplished by Rabi driving a qubit and applying sideband tones to a coupled resonator mode. Notably, the same scheme can be applied to trapped ions by simply employing the appropriate sideband tones.

We consider a transmon qubit dispersively coupled to a high-Q mode of a superconducting resonator. We apply a Rabi drive to the qubit and four sideband drives to the resonator.
Assuming that the Rabi drive is small compared to the transmon anharmonicity, the higher levels of the transmon can be ignored and the transmon is regarded as a qubit. In the frame rotating with the Rabi drive frequency and with the resonator frequency the Hamiltonian takes the form~\cite{git}
\begin{equation}
\label{eq:dispersive}
    \begin{split}
           H &= H_\mathrm{q}+H_\mathrm{drive} +H_\mathrm{disp} \\
           H_\mathrm{q} &=\frac{\Omega_R}{2}\sigma_x  + \frac{\Delta_q}{2}\sigma_z \\ H_\mathrm{drive}&=\epsilon(t)a+\epsilon(t)^*a^\dagger \\
           H_\mathrm{disp}&= \frac{\chi}{2}\sigma_z a^\dagger a, 
    \end{split}
\end{equation}
where $\Omega_R$ is the Rabi frequency, $\Delta_q$ is a small detuning frequency, $\chi$ is the dispersive shift, and $\epsilon(t)$ is the time dependent drive of the resonator. We choose $\epsilon(t)$ to describe two pairs of symmetric sidebands detuned from the resonator frequency by $\pm\Omega_R\pm\delta\Omega$ as depicted in Figure~\ref{fig:schemeOverview}(b) and given by
\begin{equation}
\begin{split}
    \epsilon(t) &= \bar{a}_0\left[ (\Omega_R-\delta\Omega)\sin((\Omega_R-\delta\Omega)t)\right. \\&-\left.\mathrm{i}(\Omega_R+\delta\Omega)\cos((\Omega_R+\delta\Omega)t)  \right],
\end{split}
\end{equation}
where $\bar{a}_0$ is a complex displacement amplitude. This results in a periodic displacement of the cavity mode in phase space, according to
\begin{equation}
    \begin{split}
        \bar{a}(t)=\bar{a}_0\left(\sin((\Omega_R+\delta\Omega)t)-\mathrm{i}\cos((\Omega_R-\delta\Omega)t)\right),
    \end{split}
\end{equation}
where we assume an initial state of $-\mathrm{i}\bar{a}_0$. This can be achieved by applying a separate displacement pulse or shaping the waveform of the sidebands pulse accordingly. 
By performing a displacement transformation $U(t)=D(\bar{a}(t))=\exp(\bar{a}(t)a^\dagger-\bar{a}(t)^*a)$, $H_\mathrm{drive}$ is eliminated and $a\rightarrow a-\bar{a}(t)$, such that the interaction becomes
\begin{equation}\label{eq:Hdisplaced}
\begin{split}
    H_\mathrm{disp}'&= \frac{\chi}{2}\sigma_z\left[a^\dagger a -\bar{a}(t)a^\dagger-\bar{a}(t)^*a\right. \\ &+ \left.|\bar{a}_0|^2 \left(1+\sin(2\delta\Omega t)\sin(2\Omega_Rt)\right) \right].
\end{split}
\end{equation}
We neglect the last term $\chi\sigma_z|\bar{a}_0|^2 \sin(2\delta\Omega t)\sin(2\Omega_Rt)/2$
due to the rotating wave approximation (RWA), under the assumption that $\Omega_R\gg\chi|\bar{a}_0|^2$, and eliminate the AC Stark shift term $\chi|\bar{a}_0|^2\sigma_z/2$ by setting $\Delta_q=-\chi|\bar{a}_0|^2/2$.

We now go into the Hadamard frame by renaming $\sigma_x\leftrightarrow\sigma_z$ and then go into the frame that oscillates at the Rabi frequency by applying $U(t)=\exp\left(\mathrm{i}\sigma_z\Omega_Rt/2\right)$. These transformations eliminate $H_\mathrm{q}$, and map $\sigma_z$ to $\sigma_+e^{\mathrm{i\Omega_Rt}}+\sigma_-e^{-\mathrm{i\Omega_Rt}}$. By using the RWA again to neglect all the terms that oscillate at or faster than $\Omega_R-\delta\Omega$, we arrive at the modulated coupling of Equation~\ref{eq:Hcoupling}~\cite{git}. The effective coupling strength is $g=\chi\bar{a}_0/2$, leading to a squeezing rate of $\chi^2\bar{a}_0^2/8\delta\Omega$.

\subsection{Two-mode squeezing Hamiltonian}
A very similar scheme can be applied to achieve conditional two-mode squeezing. The main difference is that now the qubit interacts via JC interaction with the first resonator and via anti-JC with the second resonator. In the experimental implementation this means that the two upper sidebands are applied to one mode and the two lower sidebands are applied to the other mode. First, we consider the modulated coupling with two modes
\begin{equation}\label{eq:HcouplingTMS}
\begin{split}
        H_\mathrm{coup}^\mathrm{TMS} &=\mathrm{i}\cos(\delta\Omega t)\left(g\sigma_+a - g^*\sigma_-a^\dagger\right) \\
        &-\sin(\delta\Omega t) \left(g\sigma_+b^\dagger+g^*\sigma_-b\right),
\end{split}
\end{equation}
where $a$ and $b$ are the annihilation operators of resonators $\mathrm{A}$ and $\mathrm{B}$, respectively.

Again, we use the Magnus expansion to approximate the mean Hamiltonian over a time of $2\pi/\delta\Omega$~\cite{git}. We obtain the Hamiltonian
\begin{equation}\label{eq:HsqueezingTMS}
    H_\mathrm{squeezing} = \frac{1}{2\delta\Omega}\sigma_z\left(\left(g^*\right)^2 a b + g^2 a^\dagger b^\dagger \right) + \mathcal{O}\left(\left(\frac{g}{\delta\Omega}\right)^2\right),
\end{equation}
which now describes conditional two-mode squeezing.

We now consider a system composed of a transmon qubit dispersively coupled to two resonators or two modes of the same superconducting cavity. The driven system is described by
\begin{equation}
    \begin{split}
           H^\mathrm{TMS} &= H_\mathrm{q}^\mathrm{TMS}+H_\mathrm{drive}^\mathrm{TMS} +H_\mathrm{disp}^\mathrm{TMS} \\
           H_\mathrm{q}^\mathrm{TMS} &=\frac{\Omega_R}{2}\sigma_x  + \frac{\Delta_q}{2}\sigma_z \\ H_\mathrm{drive}^\mathrm{TMS}&=\epsilon_\mathrm{A}(t)a+\epsilon_\mathrm{A}(t)^*a^\dagger +\epsilon_\mathrm{B}(t)b+\epsilon_\mathrm{B}(t)^*b^\dagger\\
           H_\mathrm{disp}^\mathrm{TMS}&= \frac{\chi_\mathrm{A}}{2}\sigma_z a^\dagger a + \frac{\chi_\mathrm{B}}{2}\sigma_z b^\dagger b, 
    \end{split}
\end{equation}
The drives are now separated such that the lower sidebands are applied to the first mode and the upper sidebands are applied to the seconds mode.

This is given by
\begin{equation}
\begin{split}
    \epsilon_{a,b}(t) &= \mathrm{i}\frac{\chi}{\chi_{a,b}}\bar{a}_0\left[ (\Omega_R-\delta\Omega)e^{\pm\mathrm{i}(\Omega_R-\delta\Omega)t}\right. \\&-\left.(\Omega_R+\delta\Omega)e^{\pm\mathrm{i}(\Omega_R+\delta\Omega)t}  \right],
\end{split}
\end{equation}
such that each mode is periodically displaced by
\begin{equation}
    \begin{split}
        \bar{a}(t)&=-\mathrm{i}\bar{a}_0\frac{\chi}{\chi_\mathrm{A}}\mathrm{i}e^{\mathrm{i}\Omega_Rt}\cos(\delta\Omega t) \\
        \bar{b}(t)&=\mathrm{i}\bar{a}_0\frac{\chi}{\chi_\mathrm{B}}\mathrm{i}e^{-\mathrm{i}\Omega_Rt}\sin(\delta\Omega t),
    \end{split}
\end{equation}
where we assume that at time zero mode $a$ is in coherent state $-\mathrm{i}\bar{a}_0$ and mode $b$ is in the vacuum state. This ensures that both modes are in the vacuum state in the displaced frame. To set mode $a$ in the right state we may start the drive on mode $a$ a quarter period before applying the other drives, or apply a displacement pulse.
We perform a displacement transformation on each mode, that is defined by $U(t)=D_\mathrm{A}(\bar{a}(t))D_\mathrm{B}(\bar{b}(t))$. This eliminates $H_\mathrm{drive}^\mathrm{TMS}$ and takes $a\rightarrow a -\bar{a}(t)$ and $b\rightarrow b - \bar{b}(t)$. We obtain the Hamiltonian
\begin{equation}\label{eq:HdisplacedTMS}
\begin{split}
    H'^{\mathrm{TMS}}_\mathrm{disp}& = \frac{1}{2}\sigma_z\left[\chi_\mathrm{A} a^\dagger a -\chi_\mathrm{A}\bar{a}(t)a^\dagger-\chi_\mathrm{A}\bar{a}(t)^*a \right. \\ 
    &+ \chi_\mathrm{B} b^\dagger b-\chi_\mathrm{B}\bar{b}(t)b^\dagger-\chi_\mathrm{B}\bar{b}(t)^*b \\ &+|\bar{a}_0|^2\chi^2(1/\chi_\mathrm{A}-1/\chi_\mathrm{B})\cos\left(2\delta\Omega t\right)/2 \\
    &\left.+|\bar{a}_0|^2\chi^2(1/\chi_\mathrm{A}+1/\chi_\mathrm{B})/2\right].
\end{split}
\end{equation}
We eliminate the AC Stark shift term by setting $\Delta_q=-|\bar{a}_0|^2\chi^2(1/\chi_\mathrm{A}+1/\chi_\mathrm{B})/2$, but, unlike the single mode case, we are left with another time dependent term that comes from the difference in coupling strengths  $\chi_\mathrm{A}-\chi_\mathrm{B}$. This term will soon be eliminated by the RWA after the following transformation. We go to the Hadamard frame and the frame rotating at the Rabi frequency, and neglect all the terms that oscillate at or faster than $\Omega_R-\delta\Omega$, to reach the modulated coupling of Equation~\ref{eq:HcouplingTMS}~\cite{git}.

\subsection{Squeezing limitations}
We wish to quantify the maximum squeezing amplitude before our approximations break down and further squeezing is impeded. We will give a rough estimate for the maximum photon number that can be stored in the generated squeezed state. 

The RWA that was used to approximate the full driven Hamiltonian of Equation~\ref{eq:Hdisplaced} to get the modulated coupling of Equation~\ref{eq:Hcoupling}, can only be applied under several assumptions. Some of the assumptions regard dynamic values that depend on the state of the system at any given time. Mainly, the number of photons in the resonator. We demand that $\Omega_R\gg \chi n /2$ and $\Omega_R\gg\chi|\bar{a}_0|\sqrt{n}/2$, where $n$ is the photon number. Therefore, our squeezed state is limited in size by roughly $n\leq 2\Omega_R/\chi$ and $n\leq (2\Omega_R/\chi|\bar{a}_0|)^2$, before high order terms come into effect. As the Rabi frequency $\Omega_R$ must be smaller than the anharmonicity of the transmon to prevent population of higher excited states, we must limit $\chi$ and $\chi \bar{a}_0$. Naively, it seems one can minimize $\chi$ while keeping $\chi\bar{a}_0$ constant by increasing the sideband amplitudes $\bar{a}_0$. However, experimental restriction limit the sideband drive amplitudes to about $400$ MHz~\cite{Eickbusch2022}, so that $\bar{a}_0$ is limited to about $400 \mathrm{MHz}\cdot 2 \pi/\Omega_R$.

Another limitation on the photon number of the squeezed state comes from the estimation of the conditional squeezing Hamiltonian using the Magnus expansion (Equation~\ref{eq:Hsqueezing} and Equation~\ref{eq:HsqueezingTMS}). For the approximation to hold, the interaction strength $\chi\bar{a}_0\sqrt{n}/2$ must be smaller than the detuning $
\delta\Omega$. This roughly sets the limit on the photon number as $n\leq(2\delta\Omega/\chi\bar{a}_0)^2$. When these limits are reached we must consider the next order term in $\bar{a}_0\chi/\delta\Omega$~\cite{git}, that is given by
\begin{equation}
    H_\mathrm{2} = \frac{\mathrm{i}\chi^3}{8\delta\Omega^2} \sigma_-\left(\bar{a}_0^3\left(a^\dagger\right)^3+\bar{a}_0|\bar{a}_0|^2\left(a+a^\dagger a^2\right) \right) + \mathrm{H.C.}
    \label{eq:HighOrderTerm}
\end{equation}

% \begin{figure}[ht]
%     \centering
%     \includegraphics[width=0.8\linewidth]{Fig1.\gfxext}
%     \caption{Drive frequencies around cavity modes for generation of single-mode squeezing \figCap{a} and two-mode squeezing \figCap{b}. The linewidths of the cavity modes and frequency detunings are not to scale. .}
%     \label{fig:Freqs}
% \end{figure}

 \begin{figure*}[t]
    \centering
    \begin{overpic}[width=1\linewidth]{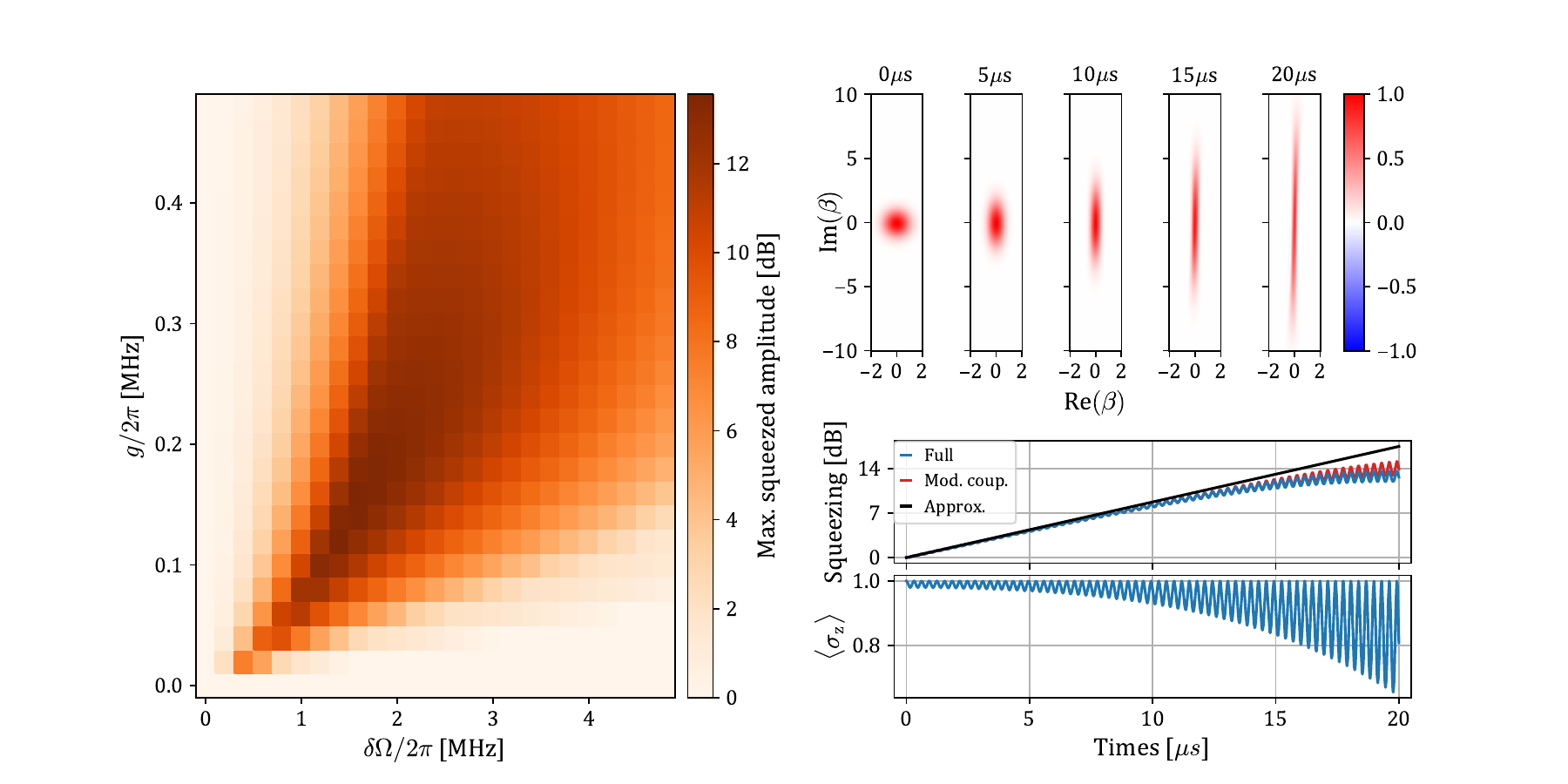}
        \put(8.1,43.8){\figCap[0]{a}}
        \put(51,43.8){\figCap[0]{b}}
        \put(51,22.5){\figCap[0]{c}}
    \end{overpic}
    \caption{\textbf{Conditional single-mode squeezing.} The squeezing of the $X$ (blue) or $Y$ (red) quadratures post-selected on the qubit state being $|g\rangle$ or $|e\rangle$, respectively. The black solid line represents the approximated conditional squeezing of Eq.~\ref{eq:Hsqueezing} and the black dashed line represents the dynamics of the same Hamiltonian but with the added higher order terms from Eq.~\ref{eq:HighOrderTerm}. The inset shows the fast oscillations between squeezing and anti-squeezing of the two quadratures conditioned on the state of the qubit. The oscillations are anti-phase locked, making it so both quadratures are never squeezed together, except for small squeezing amplitudes.}
    \label{fig:Fig2}
\end{figure*}

\section{Results and discussion}

We simulated the full dynamics in the displaced frame, as captured by $H_\mathrm{q} + H'_\mathrm{disp}$ for single-mode squeezing (Equation~\ref{eq:Hdisplaced}) and $H_\mathrm{q} + H'^{\mathrm{TMS}}_\mathrm{disp}$ for two-mode squeezing (Equation~\ref{eq:HdisplacedTMS}) (the simulations code is available at~\cite{git}). This allowed us to keep the Hilbert space of the resonator small relative to the undisplaced frame, without applying any approximations. We also simulated the modulated coupling of Equation~\ref{eq:Hcoupling} for single-mode squeezing and Equation~\ref{eq:HcouplingTMS} for two-mode squeezing. In addition, we simulated the conditional squeezing Hamiltonian of Equation~\ref{eq:Hsqueezing} for single-mode squeezing and Equation~\ref{eq:HsqueezingTMS} for two-mode squeezing.

The simulation results should be reproducible in an experiment on current systems comprising a long-lived harmonic oscillator dispersively coupled to a qubit. One may measure the Wigner characteristic function of a single mode~\cite{Eickbusch2022}, or the joint-Wigner characteristic function of two modes~\cite{diringerConditionalnotDisplacementFast2024a}.

\subsection{Preparation of single-mode squeezed vacuum}

Figure~\ref{fig:Fig2}~(a) shows a two-dimensional sweep of the maximum squeezing over $\delta\Omega$ and $g$ under the full evolution of the driven system. We set $\chi/2\pi=-50$ kHz, which is usual for a memory mode in circuit QED~\cite{Eickbusch2022, malul2023}, and $\Omega_R/2\pi=40$ MHz, which can be realized experimentally~\cite{Hacohen-Gourgy2016}. Figure~\ref{fig:Fig2}~(c) shows the time evolution of the squeezed state and the state of the qubit, for the optimal squeezing parameters of $g/2\pi=0.16$ MHz and $\delta\Omega/2\pi=1.6$ MHz found in Figure~\ref{fig:Fig2}~(a). The maximum achieved squeezing is about 13.5 dB, slightly lower than the highest demonstrated intra-cavity squeezing in circuit QED~\cite{Cai2025Quantum}. However, our setup does not require Kerr nonlinearity, which leads to state distortion~\cite{Eickbusch2022}. The highest demonstrated intra-cavity squeezing in a negligibly weak Kerr nonlinearity setup was $11.1$ dB~\cite{Eickbusch2022}. As we did not consider the effects of decoherence, achieving the simulated result in an experiment would be challenging, but not impossible. The squeezing was generated after about $20~ \mu s$, shorter than the lifetimes of current state-of-the-art transmon-cavity setups by an order of magnitude~\cite{Sivak2023Real-time}. Furthermore, the Rabi drive in this squeezing scheme provides dynamic decoupling of the qubit from low frequency noise sources and therefore increases the dephasing time.

\subsection{Preparation of superposition of squeezed states}
\begin{figure}[ht]
    \centering
    \includegraphics[width=1\linewidth]{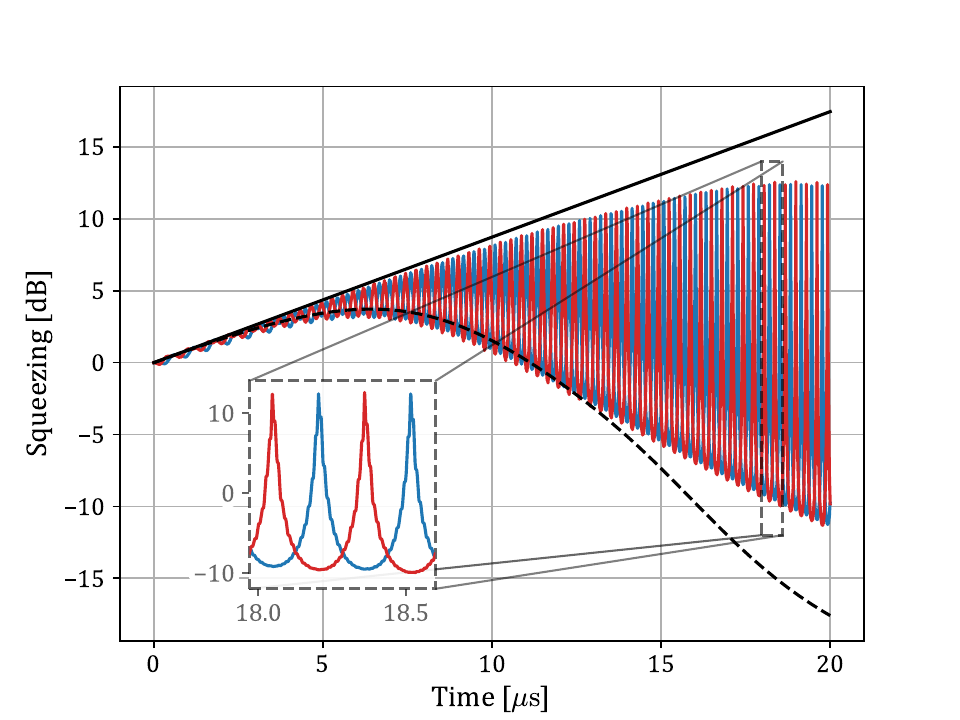}
    \caption{\textbf{Conditional single-mode squeezing.} The squeezing of the $X$ (blue) or $Y$ (red) quadratures post-selected on the qubit state being $|g\rangle$ or $|e\rangle$, respectively. The black solid line represents the approximated conditional squeezing of Eq.~\ref{eq:Hsqueezing} and the black dashed line represents the dynamics of the same Hamiltonian but with the added higher order terms from Eq.~\ref{eq:HighOrderTerm}. The inset shows the fast oscillations between squeezing and anti-squeezing of the two quadratures conditioned on the state of the qubit. The oscillations are anti-phase locked, making it so both quadratures are never squeezed together, except for small squeezing amplitudes.}
    \label{fig:condSqueeze}
\end{figure}

We simulated evolution of the fully driven system with the optimal parameters from Figure~\ref{fig:Fig2} (a) when the qubit is initialized in $|+\rangle = \left(\ket{g}+\ket{e}\right)/\sqrt 2$. The idea here is to show the conditional nature of the operation. Specifically, to generate entanglement between the qubit and the oscillator such that we will end up with a state of the form \begin{equation}
    \ket{\psi_\mathrm{entangled}}=\mathrm{CS}(\xi)\ket{+,0}=\frac{1}{N(\xi)}\left(\ket{e,\xi}+\ket{g,-\xi}\right),
\end{equation}
where the conditional squeezing operation is defined by $CS(\xi) = S(\sigma_z \xi)=\ket{e}\bra{e}\otimes S(\xi)+\ket{g}\bra{g}\otimes S(-\xi)$, $S(\xi)$ is the squeezing operator and $N(\xi)$ is a normalization factor. Figure~\ref{fig:condSqueeze} shows that the state of the composite system does become entangled and the squeezing direction is clearly conditioned on the state of the qubit. However, the higher order effects, captured in Equation~\ref{eq:HighOrderTerm}, quickly interfere with the conditional squeezing, thus limiting the amplitude of the superposition of entangled squeezed states to about $4~\mathrm{dB}$.

\subsection{Preparation of two-mode squeezed vacuum}
\begin{figure}[ht]
    \centering
    \begin{overpic}[width=1\linewidth]{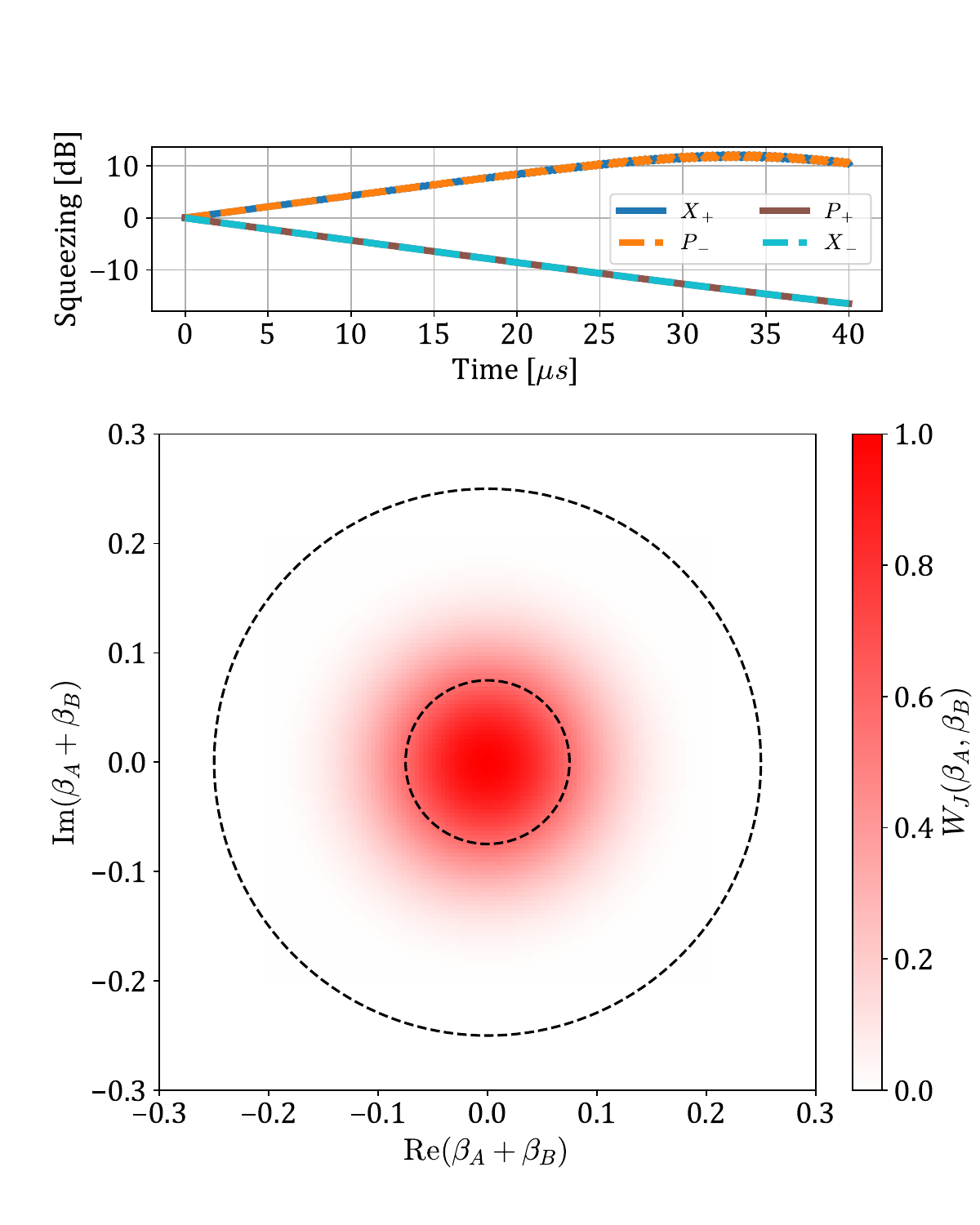}
        \put(1.5,90){\figCap[0]{a}}
        \put(1.5,66){\figCap[0]{b}}
    \end{overpic}
    \caption{\textbf{Generation of two mode squeezed vacuum.} 
    \figCap{a} The time evolution of the squeezing amplitude of the sum and difference of quadratures of the two modes. 
    \figCap{b} A 2D cut of the joint-Wigner function of the two-mode squeezed state for the maximal squeezed state with $12.1~\mathrm{dB}$ of squeezing, achieved after about $33~\mathrm{\mu s}$. The dashed black circles represent the variance of the squeezed state (small) and the variance of a vacuum state (large).}
    \label{fig:TMSV}
\end{figure}

Figure~\ref{fig:TMSV} shows the dynamics of a two-mode squeezed vacuum evolving under our continuous squeezing scheme with the qubit initialized in $\ket{g}$. The characteristic simultaneous squeezing of the $X_+=X_a+X_b=\left(a^\dagger+a+b^\dagger+b\right)/2$ and $P_+=P_a+P_b=\mathrm{i}\left(a^\dagger-a+b^\dagger-b\right)/2$ quadratures is clearly observed in the simulation.

\section{Universal control using controlled-squeezing}

Universal control of a system is the promise that repeated applications of Hamiltonians from a given set can generate all unitary operations from within the Hilbert space of said system \cite{lloyd1995almost}. 
To learn what operations can be effectively generated by repeatedly applying controllable Hamiltonians, one needs to explore the span of operators created by commutations-relations and linear combinations of existing operators \cite{lloyd1995almost}.
Universal control of the quantum harmonic oscillator cannot be achieved with only displacement and squeezing operations \cite{lloydQuantumComputationContinuous1999}, but here we show that with the addition of our suggested controlled-squeezing operation, full control can be achieved, allowing the creation of all Bosonic states.

The commutator of conditional-squeezing %
$
    (p^2 - q^2) \sigma_z
$ %
and unconditional-squeezing %
$
     p^2
$
is a new conditional-squeezing operator
$
    \left[ 
        \left(
            p^2 - q^2
        \right)
        \sigma_z
        ,
        p^2
    \right]
    =
    2\sigma_z 
    -
    2i p q \sigma_z 
$. %
Combining this with basic qubit operators and scaling, yields the Hamiltonian %
$
    p q \sigma_z 
$. % 
Interestingly, commuting this with simple displacement operators results in effective controlled-displacement \cite{Eickbusch2022, diringerConditionalnotDisplacementFast2024a} operators %
$
    \left[
        p q \sigma_z 
        ,
        q
    \right]
    =
    -i q \sigma_z
$ %
and %
$
    \left[
        p q \sigma_z 
        ,
        p
    \right]
    =
    i p \sigma_z
$. %
By rotating the qubit before and after controlled-operations, we effectively change the qubit's axis through which the operation is determined:
$   
    \left[
        (p^2 - q^2) \sigma_z 
        ,
        \sigma_x
    \right]
    =
    i (p^2 - q^2) \sigma_y
$ %
and %
$
    \left[
        q \sigma_z 
        ,
        \sigma_x
    \right]
    =
    i q \sigma_y
$. %

While $q$ and $p$ operators to the $2^{\text{nd}}$ power cannot provide higher polynomials by themselves \cite{lloydQuantumComputationContinuous1999}, the conditioned-operators can. For example: 
$
    \left[
        q \sigma_x 
        ,
        q \sigma_y
    \right]
    =
    i q^2 \sigma_z
$ %
and %
$
    \left[
        q^2 \sigma_x 
        ,
        q \sigma_y
    \right]
    =
    i q^3 \sigma_z
$. %
In the same way, richer polynomials can be generated like
$
    \left[
        q^3 \sigma_x 
        ,
        p \sigma_y
    \right]
    =
    i
    q^3 
    p 
    \sigma_z
    +
    i
    p 
    q^3 
    \sigma_z 
$. %
Operators that act only on the harmonic-oscillator can be achieved, as is evident by deriving commutation-relations between two controlled-operations. For example %
$
    \left[
        q^3 
        p 
        \sigma_z
        ,
        p \sigma_z
    \right]
    =
    3i q^2 p 
$. %

\section{Conclusions}

We have presented a novel scheme for the continuous intra-cavity generation of both single- and two-mode squeezed states using a Rabi-driven qubit dispersively coupled to quantum harmonic oscillators. By engineering a modulated Jaynes-Cummings interaction through Rabi and sideband drives, our approach enables conditional squeezing, in which the squeezed quadrature is directly linked to the qubit state. 

While our simulations predict significant squeezing levels—up to 13 dB for single-mode and 12 dB for two-mode squeezing—the amplitude of the squeezed superposition remains limited by higher-order effects. This highlights a couple of promising directions for future research: (i) alter the scheme to mitigate high-order corrections and enhance the effective squeezing, (ii) extending the method to multi-mode systems for more complex quantum state engineering. In summary, our work lays a strong foundation for exploring conditional squeezing as a versatile tool in advanced quantum architectures.
\section{Acknowledgements}
This research was partially funded by the Israeli Science Foundation (ISF), the Binational Science Foundation (BSF), and the Hellen Diller Quantum Center at Technion Israel Institute of Technology.

\bibliography{bib}

% %%%%%%%%% Merge with supplemental materials %%%%%%%%%%
% \pagebreak
% \clearpage
% \newpage
% \input{suppMat}

\end{document}